\documentclass[aps,pre,showpacs,twoside,amsfonts,twocolumn]{revtex4}
  
\usepackage{graphicx}
\usepackage{bm}

\renewcommand{\d}{\mathrm{d}}

\begin{document}

\title{Normal stresses at the gelation transition}

\author{Kurt Broderix}
\thanks{Deceased} 

\author{Peter M\"uller}

\author{Annette Zippelius}

\affiliation{Institut f\"ur Theoretische Physik,
  Georg-August-Universit\"at, D--37073 G\"ottingen, Germany}

\date{\today}

\begin{abstract}
  A simple Rouse-type model, generalised to incorporate the effects of
  chemical crosslinks, is used to obtain a theoretical prediction for
  the critical behaviour of the normal-stress coefficients $\Psi_{1}$
  and $\Psi_{2}$ at the gelation transition. While the exact
  calculation shows $\Psi_{2}\equiv 0$, a typical result for these
  types of models, an additional scaling ansatz is used to demonstrate
  that $\Psi_{1}$ diverges with a critical exponent $\ell = k+z$.
  Here, $k$ denotes the critical exponent of the shear viscosity and
  $z$ the exponent governing the divergence of the time scale in the
  Kohlrausch decay of the shear-stress relaxation function.  For
  crosslinks distributed according to mean-field percolation, this
  scaling relation yields $\ell =3$, in a accordance with an exact
  expression for the first normal-stress coefficient based on a
  replica calculation.  Alternatively, using three-dimensional
  percolation for the crosslink ensemble we find the value $\ell
  \approx 4.9$. Results on time-dependent normal-stress response are
  also presented.
\end{abstract}

\pacs{64.60.Ht, 
  61.25.Hq,     
  61.20.Lc      
  }

\maketitle

\section{Introduction}

Chemical gelation is the process of randomly introducing crosslinks
between the constitutents of a (macro-) molecular fluid. One way to
investigate the effects of the crosslinks on the fluid dynamics
consists in measuring the stresses the crosslinked fluid builds up
when subjected to a simple shear flow.  For an incompressible,
isotropic fluid one can experimentally access \cite{BiAr87} three
independent components of the stress tensor $\boldsymbol{\sigma}$: the shear
stress $\sigma_{xy}$ and the first and second normal stress
differences $\sigma_{xx}-\sigma_{yy}$ and $\sigma_{yy}-\sigma_{zz}$.
For static shear flows these give rise to three independent material
functions: the shear viscosity $\eta$ and the first and second
normal-stress coefficients $\Psi_{1}$ and $\Psi_{2}$. Generally
speaking, both Newtonian and non-Newtonian fluids possess a
non-vanishing shear viscosity. But, whereas for a Newtonian fluid both
$\Psi_{1}$ and $\Psi_{2}$ are always zero, it is precisely the
non-vanishing of $\Psi_{1}$ that explains a
number of characteristic effects known for e.g.\ polymeric liquids
\cite{Wei47}, see also \S{}2.3 in \cite{BiAr87}. On
the other hand, even for non-Newtonian fluids $\Psi_{2}$ is typically
found to be very small as compared to $\Psi_{1}$, and the
``Weissenberg hypothesis'', $\Psi_{2}=0$, is a good approximation in
these cases \cite{Wei47}. It also seems that $\Psi_{2}$ is not as well
investigated experimentally as $\Psi_{1}$.

In the context of gelation one is particularly interested in the
dependence of these stresses on the crosslink concentration $c$.
Universal critical behaviour is expected to occur at the gelation
transition, that is, at the critical concentration
$c_{\mathrm{crit}}$, where the fluid (sol) undergoes a sudden phase change
into an amorphous solid state (gel). As far as shear stress is concerned,
there exist numerous experimental investigations 
on the static shear viscosity and on the time-dependent
shear-stress relaxation function. 
The experimentally measured values for the critical exponent $k$,
which governs the algebraic divergence of the shear viscosity when
approaching $c_{\mathrm{crit}}$ from the sol side, scatter
considerably and are found in the range $k\approx 0.6
\ldots 1.7$, see
e.g.\ \cite{AdDe81,MaWi88,MaAd88,DeBo93,CoGi93,ToFa01}. The origin of
this wide spreading is controversially  
discussed and eventually unclear. From a theoretical point of view
there exists a bunch of competing and partially contradicting scaling
relations which express $k$ in terms of percolation exponents. Each of
them relies on heuristic 
arguments whose validity is mostly unclear. We refer the reader to
\cite{BrLo99,BrLo01} for a summary and references. Here we only
mention the scaling relation $k=2\nu -\beta$ which was first proposed
by de~Gennes \cite{Gen78} and rederived by many others. Erroneously,
it is generally referred to as the ``Rouse expression'' for 
the viscosity exponent. Here, $\nu$ is the exponent governing the
divergence of the correlation length and $\beta$ is associated to the
gel fraction. For three-dimensional bond percolation one would get the value 
$\left.\vphantom{\big(}(2\nu-\beta)\right|_{d=3}\approx 1.35$.
Recently, the viscosity was \emph{exactly} determined within the Rouse
model for gelation in \cite{BrLo99,BrLo01}. The analysis disproves the
above result and shows that 
\begin{equation}
  \label{phibeta}
  k= \phi -\beta
\end{equation}
is the true scaling relation valid for Rouse dynamics. Here, $\phi$
denotes the first crossover exponent of a
corresponding random resistor network \cite{HaLu87,StJaOe99}. 
When inserting \cite{BrLo99,BrLo01} high-precision data for $\phi$ and
$\beta$ obtained from three-dimensional bond
percolation, the true Rouse value of the viscosity exponent turns out to be
$\left.\vphantom{\big(}(\phi-\beta)\right|_{d=3}\approx 0.71$ 
and agrees with simulations \cite{VePl01} on a similar model.
The discrepancy to de Gennes' result above can be attributed to the
neglect of the \emph{multi}fractal nature of percolation clusters in
\cite{Gen78}. Amazingly, the true Rouse value 
$\left.\vphantom{\big(}k\right|_{d=3}\approx 0.71$
differs only little from that of another proposal, $k=s$, by de Gennes
\cite{Gen79}, where he alluded to an analogy to the conductivity exponent    
$\left.\vphantom{\big(}s\right|_{d=3}\approx 0.73$ of an electrical
network consisting of a random mixture of superconductors and normal
conductors. This close agreement, however, seems to be coincidental. 

In contrast, the authors are not aware of any experimental or
theoretical studies concerning the dependence on the crosslink
concentration $c$ of normal stresses near the gelation transition.
This seems all the more surprising since there exist many
experiments \cite{LoMe73,VeWi90,OsIn94,VeKa94} on both the shear-rate
dependence of 
normal stresses in entangled or (temporarily) crosslinked polymeric
liquids in order to explain shear-thinning or shear-thickening
phenomena and on the time dependence of the normal-stress response to
particular shapes of shear strain. Theoretical explanations of these
experimental findings mainly rely on the analysis of transient network
models, see e.g.\ \cite{TaEd92,AhOs94,AhOs95,BrHo95}.

Even though Rouse-type models incorporate no other physical
interactions between monomers apart from connectivity, they serve as 
a standard theoretical reference in terms of which
experimental data are frequently interpreted. Therefore it is
important to test their predictions as accurately as possible.  It is
the purpose of this Paper to use the same generalised Rouse-type model
as in \cite{BrLo99,BrLo01,BrAs01} to predict the critical
behaviour of the normal-stress coefficients $\Psi_{1}$ and $\Psi_{2}$
at the gelation transition. Within this model it will turn out that
$\Psi_{2}$ vanishes for all $c$ and that $\Psi_{1}$ diverges with a
critical exponent 
\begin{equation}
  \label{ellka}
  \ell = k+z
\end{equation}
when approaching $c_{\mathrm{crit}}$
from the sol side. Here, $z$ denotes the exponent governing the divergence
of the time scale in the Kohlrausch decay of the shear-stress
relaxation function.  For crosslinks distributed according to
mean-field percolation (also called ``classical theory''), this
scaling relation yields $\ell =3$, in a 
accordance with an exact expression for $\Psi_{1}$ based on a replica
calculation. Alternatively (and more realistically), using
three-dimensional percolation for the crosslink ensemble we find the
value $\ell \approx 4.9$.  Thus, the model predicts a much more
pronounced divergence of $\Psi_{1}$ as compared to $\eta$ so that
$\Psi_{1}$ may serve as a sensitive indicator for the gelation
transition. 
We also derive results on the time-dependent
normal-stress response.  In particular, the Lodge-Meissner
rule, see e.g.\ \S{}3.4.e in \cite{BiAr87}, is shown to hold for
normal-stress relaxation after a sudden shearing displacement.

We hope that these theoretical investigations motivate corresponding
experimental work in order to develop more insight on normal stresses
in (near) critical gels.

%
\section{Model}
%

We follow a semi-microscopic approach to gelation based on a
Rouse-type model for $N$ monomers. The monomers are treated as point
particles with positions ${\mathbf{R}}_i(t)$, $i=1,\ldots,N$, in
three-dimensional space. The motion of the monomers is constrained by
$M$ randomly chosen, harmonic crosslinks which connect the pairs
$(i_e,i'_e)$, $e=1,\ldots,M$, of monomers and give rise to the
potential energy
\begin{equation} \label{poten}
  U := \frac{3}{2a^2}\:\sum_{e=1}^M \lambda_e
  \bigl( {\mathbf{R}}_{i_e}-{\mathbf{R}}_{i'_e} \bigr)^2
  =: \frac{3}{2a^2}\:\sum_{i,j}^N \Gamma_{ij}\,
  {\mathbf{R}}_{i}\cdot{\mathbf{R}}_{j} \,.
\end{equation}
Here, the fixed length $a>0$ models the overall inverse coupling
strength, whereas the individual coupling constants $\lambda_{e}$ are
chosen at random. Quite often, only the special case $\lambda_{e}=1$
has been considered previously. The second equality in (\ref{poten})
introduces the random $N\times N$-connectivity matrix, which encodes
all properties of a given crosslink realisation.

\begin{figure}[t]
  \includegraphics[width=10.5pc]{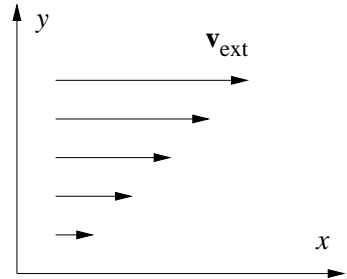}
  \caption{Homogeneous linear shear flow (\protect{\ref{flowfield}})}
  \label{shearflow}
\end{figure}

Following \cite{DoEd85,BiCu87,SoVi95} we employ a simple
relaxational dynamics
\begin{equation} \label{eqmotion}
 \zeta \left[ \partial_t R^{\alpha}_i(t)
- v^{\alpha}_{\mathrm{ext}}\bigl({\mathbf{R}}_i(t),t\bigr)\right] = 
  - \frac{\partial U}{\partial R^{\alpha}_i}(t) 
  + \xi^{\alpha}_i(t)
\end{equation}
without inertial term to describe the motion of the monomers in the
externally applied velocity field
\begin{equation}
  \label{flowfield}
  v^{\alpha}_{\mathrm{ext}}({\mathbf{r}},t) := \delta_{\alpha x}\,\kappa(t)\, y
\end{equation}
with a time-dependent shear rate $\kappa(t)$, see also
Fig.~\ref{shearflow}.  Here, Greek indices label Cartesian coordinates
$x$,$y$ or $z$.  A friction force with friction constant $\zeta$
applies if the velocity of a monomer deviates from the externally
applied flow field. The crosslinks exert a force $-\partial
U/\partial{\mathbf{R}}_i$ on the monomers, in addition to a random,
fluctuating thermal-noise force obeying Gaussian statistics with zero
mean and covariance 
$\langle\xi_{i}^{\alpha}(t)\xi_{j}^{\beta}(t')\rangle = 2\zeta
\delta_{\alpha\beta}\delta_{ij}\delta(t-t')$. Note that we have chosen
units in which the inverse temperature is equal to one.  Given the
shear flow (\ref{flowfield}), the equation of motion (\ref{eqmotion})
is linear and can be solved exactly for each realisation of the
thermal noise \cite{BrLo01}.

To complete the description of the dynamic model, we have to specify
the statistical ensemble which determines the realisations of the
crosslinks. We will distinguish two cases:
\begin{enumerate}
\item[(i)] Mean-field percolation (also called ``classical theory''):
  each pair of 
  monomers is chosen independently with equal probability $M/N^{2}$,
  irrespectively of the monomer positions in space. As a function of
  the crosslink concentration $c:=M/N$ , the system undergoes a
  percolation transition at a critical concentration
  $c_{\mathrm{crit}}= \frac12$.  For $c<c_{\mathrm{crit}}$ there is no
  macroscopic cluster, and almost all clusters are trees \cite{ErRe60,Bol98}.
\item[(ii)] Three-dimensional bond percolation \cite{StAh94,BuHa96}.
\end{enumerate}
For either case we assume the random coupling constants $\lambda_{e}$
to be distributed independently of the crosslink configuration, as
well as independently of each other with the same (smooth) probability
distribution $p(\lambda)$. Moreover, sufficiently high inverse moments
\begin{equation}
  \label{moments}
  P_{n}:= \int_{0}^{\infty}\!\d\lambda\, \lambda^{-n} p(\lambda)
\end{equation}
of $\lambda_{e}$ shall exist.


The combined average over crosslink configurations and random coupling
constants will be denoted by an overbar
$\overline{\phantom{l}{\scriptstyle\bullet}\phantom{l}}$. Using this
notation, we implicitly assume that the macroscopic limit
$N\to\infty$, $M\to\infty$, $M/N\to c$ is carried out, too.

Before turning to the analysis of the model we would like to comment
on the fact that it describes the random crosslinking of single
monomers rather than of pre-formed polymers. However, this does not
mean that the applicability of the model is limited to the
description of random networks built up by polycondensation from small
structural units. Indeed, we expect from universality that details at
small length scales are irrelevant for the true critical behaviour at the
gelation transition so that these results will also hold for random
networks built from arbitrary macromolecules, as is the case in
vulcanization, for example. This general universality argument was
confirmed \cite{BrLo01} by explicit computations of the critical
behaviour of the shear viscosity with the mean-field distribution of
crosslinks.

%
\section{Stress tensor and normal-stress coefficients}
%

Due to the externally applied shear flow the crosslinks exert shear
stress on the polymer system, whose tensor components are given in
terms of a force-position correlation \cite{DoEd85,BiCu87}
\begin{equation}
  \label{stressdef}
  \sigma_{\alpha\beta}(t) = \lim_{t_{0}\to -\infty}
  \frac{\rho_{0}}{N}\,\sum_{i=1}^{N} \left\langle\frac{\partial U}{\partial
    R_{i}^{\alpha}}(t) \; R_{i}^{\beta}(t)\right\rangle\,.
\end{equation}
Here, $\rho_{0}$ denotes the density of monomers. In (\ref{stressdef})
one has to insert the explicitly known \cite{BrLo01} solutions
${\mathbf{R}}_{i}(t)$ of 
the Rouse equation (\ref{eqmotion}) at time $t$ with initial data
${\mathbf{R}}_{i}(t_{0})$ at time $t_{0}$. In order to ensure that the
thermal-noise average allows for the description of a possible 
stationary state of the system at finite times $t$, the time evolution
is chosen to start in the infinite past, $t_{0}\to -\infty$, thereby
losing all transient effects which stem from the initial data.
This yields \cite{BrLo01} for the stress tensor
\begin{widetext}
\begin{equation}
  \label{stresstensor}
  \boldsymbol{\sigma} (t)  = \chi(0){\mathbf{1}} 
  + \int_{-\infty}^{t}\!\!\d t' \, \chi(t-t')\,
  \kappa(t') \left(\begin{array}{cc@{\qquad}c}
      2\int_{t'}^{t}\!\d s\,\kappa(s) & 1 & 0 \\
      1 & 0 & 0\\ 0 & 0 & 0\end{array}\right), 
\end{equation}
\end{widetext}
where $\mathbf{1}$ denotes the $3\times 3$-unit matrix and the stress
relaxation function is given by 
\begin{equation}
\label{stressrelax}
\chi(t):=
\frac{\rho_{0}}{N}\; {\mathrm{Tr}} \left[ (1-E_0) \,
\exp\left(- \frac{6t}{\zeta a^2} \Gamma\right)\right]\,.
\end{equation}
The symbol ${\mathrm{Tr}}$ in (\ref{stressrelax}) stands for the trace
over $N\times N$-matrices, and $E_0$ denotes the projector on the
space of zero eigenvalues of $\Gamma$, which correspond to translations
of whole clusters. The associated eigenvectors are constant within each
cluster and zero outside \cite{BrGo97,BrLo01}.  Within the simple
Rouse model the zero eigenvalues do not contribute to shear relaxation
because there is no force acting between different clusters. The only
contribution to stress relaxation is due to deformations of the
clusters.

For a time-independent shear rate $\kappa(t) \equiv \kappa$ it is
customary to define a first and second normal-stress coefficient by
\begin{equation}
  \Psi_{1} := \frac{\sigma_{xx} -
    \sigma_{yy}}{\rho_{0}\,\kappa^{2}}\,,\qquad\quad 
  \Psi_{2} := \frac{\sigma_{yy} -
    \sigma_{zz}}{\rho_{0}\,\kappa^{2}}\,.
\end{equation}
One deduces immediately from (\ref{stresstensor}) that
\begin{equation}
 \Psi_{2}=0\,,
\end{equation}
a characteristic result for Rouse-type models. In contrast, the first
normal-stress coefficient $\Psi_{1}$ is non-zero
\begin{equation}
  \label{psi1}
  \Psi_{1} = \frac{1}{2}\,\Bigl(\frac{\zeta a^{2}}{3}\Bigr)^{2}
  \frac{1}{N} \; \mathrm{Tr}\,\Bigl(\frac{1-E_{0}}{\Gamma^{2}}\Bigr)
\end{equation}
and independent of the shear rate $\kappa$.

For a macroscopic system $\Psi_{1}$ is expected to be a self-averaging
quantity. Therefore we will calculate the disorder average
$\overline{\phantom{l}{\scriptstyle\bullet}\phantom{l}}$ of
(\ref{psi1}) over all crosslink realisations and all crosslink
strengths. To do so it is convenient to introduce the averaged density
\begin{equation}
  \label{dos}
  D(\gamma): =\overline{\frac{1}{N}\,\mathrm{Tr}\,
    \bigl[(1-E_0)\,\delta(\gamma -\Gamma)\bigr]} 
\end{equation}
of non-zero eigenvalues of $\Gamma$. Physically, $D$ describes the
distribution of relaxation rates in the network (in units of $\zeta
a^{2}/6$), as is evident from the representation
\begin{equation}
 \label{laplace}
  \overline{\chi}(t) = \rho_{0} \int_{0}^{\infty}\!\d\gamma\,
  \exp\Bigl\{-\frac{6\,t\gamma}{\zeta a^{2}}\Bigr\} D(\gamma)
\end{equation}
of the disorder average of the stress relaxation function
(\ref{stressrelax}).  Various properties of the eigenvalue density $D$
are discussed in detail in \cite{BrAs01}. The average
$\overline{\Psi}_{1}$ now appears as the second inverse moment of $D$,
\begin{equation}
  \label{psi1av}
  \overline{\Psi}_{1} = \frac{1}{2}\,\Bigl(\frac{\zeta
    a^{2}}{3}\Bigr)^{2} \int_{0}^{\infty}\!\d\gamma\,
  \frac{D(\gamma)}{\gamma^{2}}\,,
\end{equation}
while the disorder-averaged
static shear viscosity $\overline{\eta} :=
\overline{\sigma}_{xy}/(\rho_{0}\,\kappa)$ is determined
\cite{BrLo99,BrLo01} by the first inverse moment
\begin{equation}
  \label{viscosity}
  \overline{\eta} =
  \frac{1}{\rho_{0}} \int_{0}^{\infty}\!\d t \,\overline{\chi}(t) =
  \frac{\zeta a^{2}}{6}
  \int_{0}^{\infty}\!\d\gamma\; \frac{D(\gamma)}{\gamma}\,.
\end{equation}

At this point one can already see that $\overline{\Psi}_{1}$ serves as
a sensitive indicator for the gelation transition. Indeed, the Jensen
inequality \cite{Bau96} implies
\begin{equation}
  \overline{\Psi}_{1} \ge \frac{2\;
    \overline{\eta}^{\;2}}{\int_{0}^{\infty}\!\d\gamma\, 
    D(\gamma)}
  \ge 2 \;\overline{\eta}^{\;2}\,,
\end{equation}
and hence
\begin{equation}
  \label{jensen}
  \ell \ge 2k
\end{equation}
with $\ell$, respectively $k$, denoting the critical exponent of
$\overline{\Psi}_{1} \sim (c_{\mathrm{crit}} -c)^{-\ell}$,
respectively $\overline{\eta}\sim (c_{\mathrm{crit}} -c)^{-k}$. 

In the two following sections we will determine the precise Rouse value of
$\ell$ for the two different types of crosslink ensembles described
above.

%
\section{Mean-field percolation}
%

\begin{figure}[t]
  \hspace*{1.4cm}\includegraphics[width=17pc]{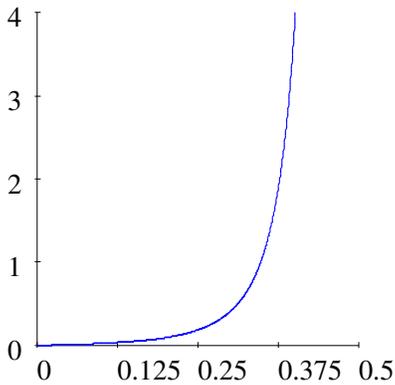}
  \caption{First normal-stress coefficient (\protect{\ref{psi1erg}})
    in units of $(\zeta a^{2}/3)^{2}$ as a function of $c$ for
    $P_{1}=P_{2}=1$.}
  \label{normalstressplot}
\end{figure}

For mean-field random graphs (i) the second inverse moment of the
eigenvalue density $D$ was calculated in Eq.\ (38) of 
\cite{BrAs01} with the help of a replica approach. This gives
rise to the exact result
\begin{widetext}
  \begin{eqnarray}
    \label{psi1erg}
    \overline{\Psi}_{1} = \frac{1}{2}\,\Bigl(\frac{\zeta
      a^{2}}{3}\Bigr)^{2} c \left[  - \frac{8c^{3} - 6c^{2} - 5c 
        + 1}{30c(1-2c)^{3}}\,P_{1}^{2}    
      - \frac{4c^{2} - 3c
        -1}{24c(1-2c)^{2}} \, P_{2} 
      + \frac{5P_{2}-4P_{1}^{2}}{240c^{2}}\,\ln(1-2c)\right]\,,
  \end{eqnarray}
\end{widetext}
which is valid for all $0<c<c_{\mathrm{crit}}=\frac{1}{2}$. The inverse
moments $P_{n}$ were defined in (\ref{moments}). From (\ref{psi1erg})
we read off the critical divergence
\begin{equation}
  \overline{\Psi}_{1} \sim \Bigl(\frac{\zeta
    a^{2}}{3}\Bigr)^{2}\, \frac{P_{1}^{2}}{240}\; \varepsilon^{-3}\,,
  \qquad  \varepsilon := c_{\mathrm{crit}} - c \downarrow 0
\end{equation}
at the gelation transition, and hence the critical exponent
\begin{equation}
  \label{mfexp}
  \ell =3\,.
\end{equation}
For $c\to 0$ one expands
\begin{equation}
  \overline{\Psi}_{1} = \Bigl(\frac{\zeta
    a^{2}}{3}\Bigr)^{2}\, \frac{P_{2}}{8} \,c + {\mathcal{O}}(c^{2})\,.
\end{equation}
Fig.~\ref{normalstressplot} displays $\overline{\Psi}_{1}$ in units of
$(\zeta a^{2}/3)^{2}$ as a function of $c$ for the special case
$P_{1}=P_{2}=1$.


It is the merit of the mean-field percolation ensemble that it allows
for a variety of exact analytical calculations. However, since the
probability for a crosslink to occur does not depend on the monomers'
positions in space, this ensemble is believed to provide a fairly
unrealistic description for three-dimensional gels. For this reason we
consider an alternative crosslink ensemble in the next section, which
has been successfully used \cite{StCo82} to explain static properties
of polymer systems.

%
\section{Three-dimensional bond percolation}
%

For this ensemble of crosslinks the second inverse moment
(\ref{psi1av}) of the 
eigenvalue density $D_{\varepsilon}$ -- note
that in this section we emphasize the dependence on $\varepsilon:=
c_{\mathrm{crit}}-c$ in the notation of various quantities -- is not
known analytically. In order 
to proceed we assume that $D_{\varepsilon}$ follows a scaling law
\begin{equation}
  \label{dscale}
  D_{\varepsilon}(\gamma) \sim \gamma^{\Delta -1}\,
  f\bigl(\gamma^{*}(\varepsilon)  /\gamma\bigr)
\end{equation}
close to the critical point and for small enough $\gamma$. It is
determined by a typical relaxation 
rate $\gamma^{*}(\varepsilon)\sim \varepsilon^{z}$, which vanishes
when approaching the critical point, and a scaling function $f(x)$
that tends to a non-zero constant for $x\to 0$ and decays faster than
any inverse polynomial for $x\to\infty$.  
In particular, this gives the power-law behaviour
$D_{\varepsilon=0}(\gamma) \sim \gamma^{\Delta-1}$ asymptotically for 
$\gamma\to 0$ at criticality, in agreement with experiments
\cite{WiCh86,ChWi87}. The measured exponent values, however, scatter
considerably, $\Delta\approx 0.4\ldots 0.8$, and seem to depend on the mass of
the crosslinked molecules \cite{WiMo97}. Note that on general grounds
the exponent $\Delta$ has to be positive, because otherwise
$D_{\varepsilon=0}(\gamma)$ would not be integrable at $\gamma=0$, in
contradiction to the definition (\ref{dos}). 
The scaling
law (\ref{dscale}) yields, via the Laplace transform (\ref{laplace}),
the scaling law 
\begin{equation}
  \overline{\chi}_{\varepsilon}(t) \sim \varepsilon^{z\Delta}\,
  g\bigl(t/\tau^{*}(\varepsilon)\bigr) 
\end{equation}
for the long-time behaviour of the stress relaxation function.
Here, the scaling function obeys $g(x) \sim x^{-\Delta}$ for $x\to 0$
and the typical relaxation time $\tau^{*}(\varepsilon) := \zeta a^{2}/[6
\gamma^{*}(\varepsilon)]\sim \varepsilon^{-z}$ diverges when
approaching the critical point. Precisely at the critical point one
finds an algebraic long-time decay $\overline{\chi}_{\varepsilon =0}(t) \sim
t^{-\Delta}$. Dynamical scaling relates $\Delta$ to $z$
and to the exponent $k$ of the shear viscosity  
\begin{equation}
  \label{expeq}
  \Delta = (z-k)/z\,,
\end{equation}
see e.g.\ \cite{WiMo97,BrAs01}. For $x\to\infty$ the scaling function $g(x)$
has to decay like a stretched exponential in order to accommodate 
the experimentally found \cite{MaAd89} Kohlrausch decay 
\begin{equation}
  \label{kohlrausch}
  \overline{\chi}_{\varepsilon >0}(t) \sim
\exp\{-[t/\tau^{*}(\varepsilon)]^{\alpha}\}
\end{equation}
of the stress-relaxation function in the sol phase away from
criticality, where $\alpha$ is a non-critical and possibly
non-universal exponent. We will return to (\ref{kohlrausch}) in the
next section.

From (\ref{psi1av}), (\ref{dscale}) and (\ref{expeq}) we deduce
$\overline{\Psi}_{1} \sim\varepsilon^{-\ell}$ for
$\varepsilon\downarrow 0$ with an exponent given by the scaling
relation
\begin{equation}
  \label{scalerel}
  \ell = k + z = k\, \frac{2-\Delta}{1-\Delta}\,. 
\end{equation}
Since $\Delta>0$, we have $z>k$ and the scaling relation
(\ref{scalerel}) is compatible with the inequality (\ref{jensen}).
Eq.\ (\ref{scalerel}) was obtained previously in \cite{WiIz94} from a
model density of relaxation times with a sharp upper cut-off. 

According to (\ref{phibeta}) the viscosity exponent for the Rouse-type
model under consideration is given by $k\approx 0.71$, when using
three-dimensional bond percolation to generate the crosslink ensemble.
Concerning $\Delta$, we are only aware of \cite{BrAs01},
where this exponent is determined for the Rouse model at hand without
any further assumptions. It was done by
numerical computations of the eigenvalue density (\ref{dos}) and yields
$\Delta\approx 0.83$. But,  
as compared to $k$, we suspect the numerical accuracy of the result to
be rather poor. Yet, using these values,
Eq.\ (\ref{scalerel}) predicts
\begin{equation}
  \ell \approx 4.9
\end{equation}
for the exponent of the first normal-stress coefficient
$\overline{\Psi}_{1}$. If, instead, one ignored the
\emph{multi}fractal structure of percolation clusters in employing the
wrong scaling relations $k=2\nu -\beta$ and $t=d\nu$, where $t$ is the
critical exponent of the elastic modulus in the gel phase, one would
arrive at the value $\Delta\approx 0.66$. This would yield the
considerably lower result $\ell\approx 2.8$. Thus, it is of importance
to improve the accuracy of the exact numerical computation of $\Delta$
within the Rouse model.

Finally, we would like to point out that for mean-field percolation
the scaling relation (\ref{scalerel}) is consistent with the exact
result presented in the preceding section. For, in this case the model
yields $k=0$ \cite{BrLo99,BrLo01}, $z=3$ and $\Delta=1$ \cite{BrAs01},
and thus (\ref{scalerel}) gives $\ell =3$ in accordance with
(\ref{mfexp}).

%
\section{Time-dependent normal-stress response}
%

First, let us focus on the normal-stress response to the inception of
a steady shear flow $\kappa(t) = \kappa_{0}\,\Theta(t)$. Here $\Theta$
denotes the Heaviside-unit-step function. For each realisation of the
crosslinks Eq.\ (\ref{stresstensor}) leads to
\begin{equation}
  \label{timenormal}
  N_{1}(t):= \sigma_{xx}(t) - \sigma_{yy}(t) = 2\,\kappa_{0}^{2}
  \int_{0}^{t}\!\d t' \, t' \chi(t')
\end{equation}
in accordance with the principle of frame invariance \cite{Lar88}.
Eqs.\ (\ref{stressrelax}) and (\ref{psi1}) then imply that for all
crosslink concentrations below $c_{\mathrm{crit}}$ the first
normal-stress difference increases towards its steady-state value like
a stretched exponential
\begin{equation}
  \overline{N}_{1}(t) = \rho_{0}\,\kappa_{0}^{2}\, \overline{\Psi}_{1} -
 2\,\kappa_{0}^{2} \int_{t}^{\infty}\!\d t' \, t'\, \overline{\chi}(t')
\end{equation}
with the same exponent $\alpha$ as the shear-relaxation function
(\ref{kohlrausch}). In contrast, for $c=c_{\mathrm{crit}}$ we deduce
from (\ref{timenormal}) the algebraic growth $\overline{N}_{1}(t) \sim
t^{2-\Delta}$ for long times, a result already known on a
more phenomenological basis \cite{WiMo97}.

Second, we consider a sudden shearing displacement $\kappa(t) =
E\,\delta(t)$, where $\delta$ denotes the Dirac-delta function. From
(\ref{stresstensor}) we infer
\begin{equation}
  \label{suddennormal}
  N_{1}(t) = \int_{0}^{\infty}\!\d t'\,
  \chi(t')\; \frac{\d}{\d t'}\,\biggl(\int_{t-t'}^{t}\!\d s\,
  \kappa(s)\biggr)^{2} 
  = E^{2}\chi(t),
\end{equation}

which, after averaging over disorder, amounts to the Kohlrausch decay
(\ref{kohlrausch}) in the long-time limit for systems below the
critical point, respectively to the algebraic decay $t^{-\Delta}$ for
$c=c_{\mathrm{crit}}$. Upon comparing (\ref{suddennormal}) to the
corresponding result $\sigma_{xy}(t)=E\,\chi(t)$ for shear stress, the
Lodge-Meissner rule $N_{1}(t)/\sigma_{xy}(t) = E$, see
e.g.\ \S{}3.4.e in \cite{BiAr87}, holds for each crosslink realisation
in this Rouse-type model. 

Third, we consider the double-step strain flow $\kappa(t)=E\,\delta(t)-
E\,\delta(t-t_{1})$ with $t>t_{1}>0$. In this case one can verify in an
analogous manner the corresponding relation
$N_{1}(t)/\sigma_{xy}(t) = -E$, which is known \cite{VeKa94} to be valid 
for class~I simple fluids.

%
\section{Outlook}
%

We hope to stimulate detailed experimental investigations on the
crosslink dependence of normal stresses close to the gelation
transition. If such experimental results were at hand, one could judge
the effects of the simplifications which underly the above Rouse-type
model, such as the neglect of the excluded-volume interaction and of
the hydrodynamic interaction.

\end{document}